# Multiple-Step Quantized Triplet STDP Implemented with Memristive Synapse

Yi-Fan Liu, *Graduate Student Member, IEEE*, Da-Wei Wang, *Member, IEEE*, Zhe-Kang Dong, *Senior Member, IEEE*, and Wen-Sheng Zhao, *Senior Member, IEEE*

*Abstract*—As an extension of the pairwise spike-timing-dependent plasticity (STDP) learning rule, the triplet STDP is provided with greater capability in characterizing the synaptic changes in the biological neural cell. In this work, a novel mixed-signal circuit scheme, called multiple-step quantized triplet STDP, is designed to provide a precise and flexible implementation of coactivation triplet STDP learning rule in memristive synapse spiking neural network. The robustness of the circuit is greatly improved through the utilization of pulse-width encoded weight modulation signals. The circuit performance is studied through the simulations which are carried out in MATLAB Simulink & Simscape, and assessment is given by comparing the results of circuits with the algorithmic approaches.

## I. Introduction

MEMRISTIVE synapse-based spiking neural networks (SNNs) have recently drawn much attention in the field of novel artificial intelligence devices, due to their low operating power consumption, high integration density, and similarity with biological neural networks in functions [1]. The pairwise spike-timing-dependent plasticity (STDP) is believed to be one of the fundamental mechanisms of synaptic plasticity in biological neural networks, and it is also used as a well-known unsupervised learning rule in SNN [2]. Accordingly, numerous studies on STDP implementation on memristive synapse neural network circuits have been carried out [3].

So far, most of the existing works have been devoted to studying the pairwise STDP, where the synaptic weight is modulated according to the timing difference between one pre-synaptic neuron (Pre) spike and one post-synaptic neuron (Post) spike [2]. Although the pairwise STDP has been proven to be capable in practice, it is reported that the classical pairwise STDP was not sufficient to characterize the synaptic changes induced by pairs of spikes with increasing repetition frequency and triplets or quadruplets of spikes [4]. To solve this issue, triplet STDP is developed as an extension of classic STDP [4]. According to [5], the existing triplet STDP rules can be clarified into two groups, and they are the suppression triplet STDP, and the coactivation and timing-dependent integration rule. Compared with the first learning rule, which requires the memristor to perform forgetting behavior, the second learning rule is comparatively more difficult to achieve using circuits as it demands more complicated numerical operation and precise weight modulation. Consequently, only a few attempts have been made to implement the second learning rule in the memristive synapse-SNN until now [6], [7].

This work was supported by the National Natural Science Foundation of China under Grants 62222401, 62101170, and 61934006, the Natural Science Foundation of Zhejiang Province under Grants LD22F040003 and LXR22F040001. (Corresponding author: *Da-Wei Wang* and *Wen-Sheng Zhao*).

All authors are with the with the Zhejiang Provincial Key Lab of Large-Scale Integrated Circuits Design, School of Electronics and Information, Hangzhou Dianzi University, Hangzhou 310018, China (e-mail: davidw.zoeq@gmail.com and wsh.zhao@gmail.com).

In this brief, a mixed-signal circuit scheme called multiple-step quantized (MSQ) triplet STDP is proposed, aiming to provide a more practical and robust implementation of coactivation and timing-dependent integration triplet STDP in memristive synapse-SNN. In our scheme, the whole circuit is driven by a global clock, and the synaptic weight programming signal is encoded into pulse width by utilizing pulse width modulation (PWM). By doing this, the flexibility, robustness, and compatibility of the circuit can be improved markedly.

The rest of this work is organized as follows. In Section II, the concept of trace and the working principle of the triplet STDP model is elaborated. Then, the working principle of the proposed circuit is demonstrated in Section III, along with an ideal pulse width encoded memristive synapse model used in the following study. In Section IV, the performance of the designed circuit is studied through simulations, and verification is carried out by comparing the simulated results with those achieved by algorithmic approaches. Finally, some conclusions are drawn in Section V.

## II. Triplet STDP

As described in [8], synaptic plasticity is induced by spikes of Pre and Post, but a single spike usually cannot cause long-term plasticity. Instead, each spike can lead to an update of the internal state variable at the synapse, known as the trace and denoted by $x(t)$. According to the nearest-spike scheme [4], the trace $x(t)$ updates to its maximum value $x_{\max}$ whenever there is a neuron spike and then decreases exponentially with a time constant of $\tau$. Such dynamic behavior can be described by

$$x(t) = x_{\max} e^{-\frac{t}{\tau}}, \text{ if } t = t^f \text{ then } x \to x_{\max} \quad (1)$$

where $t^f$ denotes the moment when the neuron spike is fired.

Fig. 1(a) shows a typical neural network in which the Pre $j$ is connected to a Post $i$ through a synapse. In the minimal triplet STDP, like the pairwise STDP, each spike from the Pre contributes to a trace $x_j(t)$ and it can be expressed as

$$x_j(t) = e^{-\frac{t}{\tau_j}}, \text{ if } t = t_j^f \text{ then } x_j \to 1 \quad (2)$$

where $t_j^f$ denotes the firing times of Pre, with $x_{\max}$ set to 1. Different from the pairwise STDP, each Post spike contributes to a fast trace $y_i^1$ and a slow trace $y_i^2$ at the synapse.

$$y_i^1(t) = e^{-\frac{t}{\tau_i^1}}, \text{if } t = t_i^f \text{ then } y_i^1 \to 1 \quad (3)$$

$$y_i^2(t) = e^{-\frac{t}{\tau_i^2}}, \text{if } t = t_i^f \text{ then } y_i^2 \to 1 \quad (4)$$

where $\tau_i^1 < \tau_i^2$, namely the falling rate of $y_i^1$ is faster than $y_i^2$, as shown in Fig. 1(b).

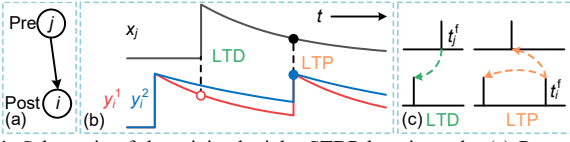

Fig. 1. Schematic of the minimal triplet STDP learning rule. (a) Presynaptic neuron $j$ is connected to a postsynaptic neuron $i$ through a synapse. (b) The spikes of presynaptic neuron $j$ contribute to a trace $x_j(t)$ and the spikes of postsynaptic neuron $i$ contribute to a fast trace $y_i^1(t)$ and a slow trace $y_i^2(t)$. (c) Spike pairing scheme to induce LTP and LTD.

As shown in Fig. 1(b) and (c), the long-term potentiation (LTP) in the triplet STDP learning rule is induced by a triplet effect. The weight increment is not only proportional to the value of the Pre trace $x_j$, but also relies on the value of the slow Post trace $y_i^2$, written as

$$\Delta W_{ij}^+(t_i^f) = A_2^+(W_{ij})x_j(t_i^f) + A_3^+(W_{ij})x_j(t_i^f)y_i^2(t_i^f - \varepsilon) \quad (5)$$

where $\varepsilon$ is a short period of time which ensures that the value of $y_i^2$ can be read just before the Post spike's arrival and Post traces' update. The long-term depression (LTD) in triplet STDP is analogous to that in classical pairwise STDP, and the weight change is proportional to the value of the fast Post trace $y_i^1$, written as

$$\Delta W_{ij}^-(t_j^f) = -A_2^-(W_{ij})y_i^1(t_j^f). \quad (6)$$

Here, $A_2^+(W_{ij})$, $A_3^+(W_{ij})$, and $A_2^-(W_{ij})$ denotes the amplitude of the weight changes. Ideally, synaptic weight change is a linear process, namely the weight change is independent of the initial weight state. In other words, for any value of $W_{ij}$, $A_2^+(W_{ij})$, $A_3^+(W_{ij})$ and $A_2^-(W_{ij})$ keep constant. In this case, $A_2^+(W_{ij})$, $A_3^+(W_{ij})$ and $A_2^-(W_{ij})$ can be substituted by constants $A_2^-$, $A_2^+$ and $A_3^+$, respectively.

## III. TRIPLET STDP IMPLEMENTATION IN MEMRISTIVE SYNAPSE-BASED NEURAL NETWORK

### A. Pulse Width-Encoded Memristive Synapse Model

Artificial synapse is one of the key components in neuromorphic circuits, and it characterizes the connection strength between two neurons through the synaptic weight, denoted by $W$. For artificial synapses implemented using memristive devices, the synaptic weights are presented by the conductance of memristors, such as in [10].

Generally, the synaptic weight can be modulated by imposing programming voltage signals across the positive and negative nodes of a memristive synapse. Ideally, the weight-updating process of artificial synapses should have linear and symmetric characteristics [11]. To achieve this, memristors with remarkable linear and symmetric LTP/LTD performance have been successfully fabricated [10]. Meanwhile, it is more practical to encode the weight modulation signal into pulse width, rather than pulse amplitude, as the pulse amplitude modulation (PAW) requires complex programmable analog circuits [12]. In this work, an ideal pulse width-encoded memristive synapse model is utilized in the simulations to ensure both rationality and simplicity. The changing speed $dW/dt$ of the synaptic weight can be calculated as

$$\frac{dW}{dt} = \begin{cases} v_P, & 0 < V_{on} \leq V \\ v_D, & V \leq V_{off} < 0 \\ 0, & V_{off} < V < V_{on} \end{cases} \quad (7)$$

where $V$ is the signal applied across the synapse, $V_{on}$ and $V_{off}$ are thresholds of LTP and LTD, $v_P$ and $v_D$ are the increasing and decreasing gradient of weight when applied the positive and negative exceeding thresholds, respectively. The variation range of synaptic weight is $W \in [W_{min}, W_{max}]$. The $v_P$ and $v_D$ are defined as constants, as the amplitude of programming signals in pulse width encoded synapses, such as the scheme in [11], is fixed.

TABLE I
THE VALUE OF PARAMETERS UTILIZED IN THE SIMULATION

| Param. | $W_{min}$ | $W_{max}$ | $V_{on}$ (V) | $V_{off}$ (V) | $v_P$ (1/s) | $v_D$ (1/s) |
|---|---|---|---|---|---|---|
| Value | 0.0 | 1.0 | 3.0 | -3.0 | 50.0 | -50.0 |

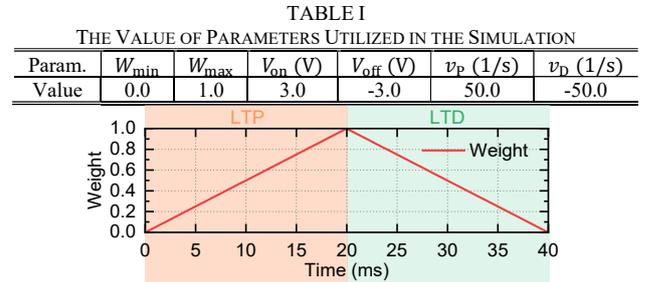

Fig. 2. (a) Potentiation and depression demonstration of the ideal memristive synapse model.

Once the amplitude is given, the increasing and decreasing gradients are determined by the sensitivity property of the synapse. The parameter values utilized in the following simulations are listed in TABLE I. Note that the models and parameters here are general or predictive only and function as generic artificial synapses. They can be substituted by other models or parameters to fit a specific application. Fig. 2(a) shows the linear LTP/LTD performance of the ideal synapse model, induced by the bi-polar stimuli with amplitudes of 4 V and -4 V.

### B. Trace Storage and Triplet STDP Implementation

The MSQ triplet STDP is proposed based on the time division multiplexing (TDM) STDP concept in [13]. The first step is to design a circuit to reproduce the trace. As described by (3), the update and decay dynamics of trace match the behaviors of a capacitor that charges to a fixed potential from a voltage source and discharges slowly through a resistor. Thus, a trace module consisting of a switch $S$, a resistor $R$, and a capacitor $C$ is designed to capture the dynamic of the race, as shown in Fig. 3(a). The trace value is represented by the capacitor voltage $V_c$, which is charged to 1 V once the $S$ is switched on by neuron spike and discharged exponentially through $R$ with a time constant of $\tau = RC$ when the $S$ is switched off, as shown in Fig. 3(b). The value of the race is sampled every millisecond, exhibited by the black dash line in Fig. 3(b). Then, the sample values are converted into pulse widths utilizing pulse width modulation (PWM), as shown in Fig. 3(c), referring to a rising-edged sawtooth carrier signal with peak amplitude $V_{Peak}$ (the blue line in Fig. 3(b)). Note that the $V_{Peak}$ is not the exact value used in the following simulations, but only for the convenience of demonstration.

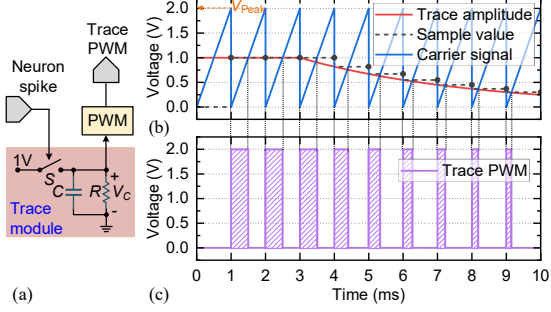

Fig. 3. (a) Block diagram of the trace module with PWM, (b) The schematic diagram of the trace, carrier signal, and sampled signal curves, and (c) The schematic of generated PWM signal.

The schematic diagram of the proposed MSQ triplet STDP circuit is depicted in Fig. 4 in which the circuit described by formulas (7) and (8) is implemented based on the circuit building blocks in MATLAB Simulink and the interconnection relationship between blocks is represented by the connecting lines. The detailed diagram schematics of the utilized circuit building blocks are exhibited in Fig. 5. In this circuit, Pre has one trace module to characterize the trace left by the Pre spike, while Post has two trace modules to characterize the fast and slow traces left by the Post spike. The slow trace of POST is delayed for 1 ms, to make its value read at the arrival of Post spike, just before updated. The delay module adopts the same design as the sample-and-hold module in PWM. For the LTP, two parameters $A_2^+$ and $A_3^+$ are utilized to control the magnitude of weight increment. Thus, amplifiers $G_1$ and $G_2$ are employed to control the relative size of $A_2^+$ and $A_3^+$. An adder and a Gilbert cell-based multiplier are utilized to implement the numerical operation. Finally, the amplitudes of weight modulation signals are converted into pulse widths through PWM modules. The peak voltages of the input sawtooth carriers of the PWM modules in Pre and Post module are $V_j^{\text{Peak}}$ and $V_i^{\text{Peak}}$, respectively, which control the magnitudes of LTP and LTD.

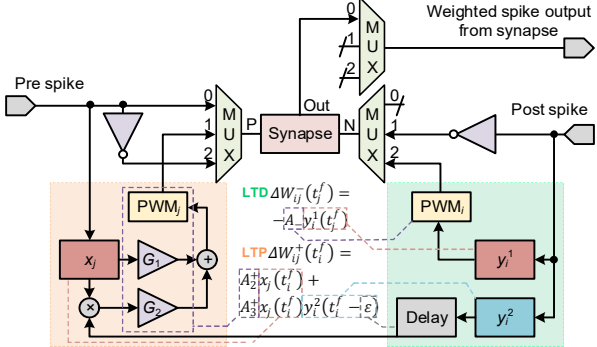

Fig. 4. Schematic circuit diagram for the proposed MSQ triplet STDP implementation with memristive synapse. Portions of the triplet STDP model and their corresponding circuit implementation modules linked by dash lines.

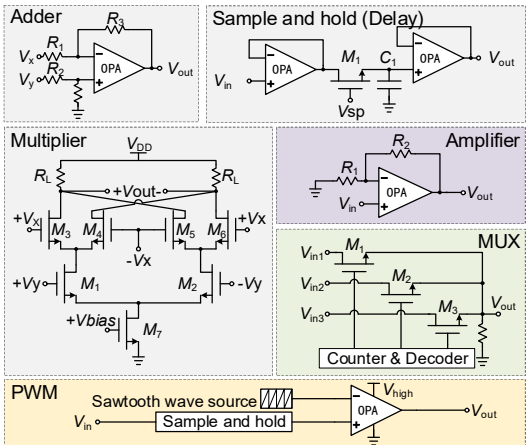

Fig. 5. Detailed diagram schematics of the utilized circuit building blocks.

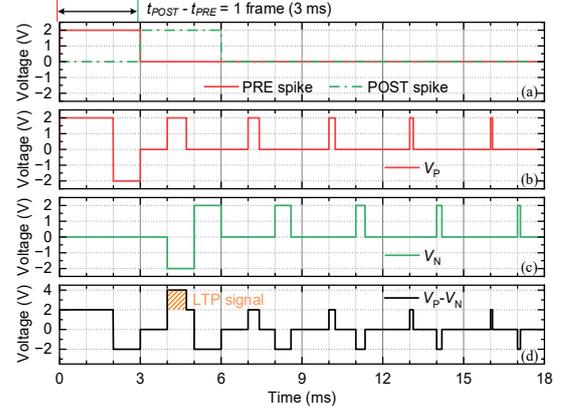

Fig. 6. Exemplary generation process of LTP signal. (a) Spikes of Pre and Post. (b) The signal sent by Pre, imposed on the positive node of the synapse. (c) The signal sent by Post, imposed on the negative node of the synapse. (d) The voltage across P/N nodes of the synapse, by over-lapping the signals on both terminals.

The TDM scheme is employed to deal with the read-write dilemma which is a typical issue of mutual interference between spike signals and weight modulation signals. The signals are transmitted by frame, with each frame comprising three timeslots. The signal imposed on the P/N nodes of the synapse is selected by the multiplexers (MUXs) in turns, which is controlled by a decoder driven by a $M = 3$ counter. An exemplary situation is illustrated in Fig. 6(a), where a Pre spike is emitted in the first frame and a Post spike is emitted in the second one. The signals imposed on the P/N nodes are shown in Fig. 6(b) and (c), respectively, and the voltage across the synapse is shown in Fig. 6(d). Thus, a LTP signal exceeding $V_{\text{on}}$ is generated, and then the LTD is by a similar way but in the following timeslot. Following, the function of the designed circuit is verified through modeling and simulating in Simulink & Simscape. To construct circuit model, the Level-1 model is adopted for CMOS devices and ideal models provided by the simulation platform are adopted for the integrated component, including the digital parts, OPA, and signal sources.

## IV. RESULTS AND DISCUSSIONS

In this section, the performance of the proposed circuit is illustrated by replicating the weight modulation results of the triplet STDP algorithm model [4], following three different protocols: pairing, triplet, and quadruplet. The parameters used in the model are divided into two groups, which are obtained by fitting the experimental results of the visual cortex and hippocampus, respectively, as listed in the upper part of TABLE II. Correspondingly, the weight modulation results of the algorithm model can be reproduced by adjusting the circuit parameters, as listed in the lower part of TABLE II. The parameter fitting procedure is as follows.

a) Set the time constants of the RC circuits in trace modules to the same value as the time constants of their corresponding trace.

b) Set $G_1$ and $G_2$ according to the relative size of $A_2^+$ and $A_3^+$.

c) Adjust $V_j^{\text{Peak}}$ and $V_i^{\text{Peak}}$ according to the magnitudes of LTP and LTD.

This work aims to implement the algorithm with a memristive circuit. Thus, the normalized mean square error (NMSE) between the circuit results and algorithmic model is used to assess the performance, calculated by

$$E = \frac{1}{P}\sum_{i=1}^{P}\left(\frac{\Delta W_{\text{cir}}^i - \Delta W_{\text{mod}}^i}{\sigma_i}\right)^2 \quad (1)$$

where $\Delta W_{\text{mod}}^i$ is the weight change obtained from the algorithmic model, and $\Delta W_{\text{cir}}^i$ is the weight change obtained from the proposed circuit, $\sigma_i$ is the measured standard error mean of $\Delta W_{\text{mod}}^i$, and $P$ is the number of measured data.

TABLE II
PARAMETERS USED IN THE MODEL AND THE PROPOSED CIRCUIT

| Algorithm | $A_2^+$ ($10^{-3}$) | $A_3^+$ ($10^{-3}$) | $A_2^-$ ($10^{-3}$) | $\tau_j$ (ms) | $\tau_i^1$ (ms) | $\tau_i^2$ (ms) |
|---|---|---|---|---|---|---|
| Visual Cortex | 0 | 50.0 | 8.0 | 16.8 | 33.7 | 40.0 |
| Hippo. Culture | 4.6 | 9.1 | 3.0 | 16.8 | 33.7 | 48.0 |
| Circuit | $V_j^{\text{Peak}}$ (V) | $[G_1, G_2]$ | $V_i^{\text{Peak}}$ (V) | $\tau_j$ (ms) | $\tau_i^1$ (ms) | $\tau_i^2$ (ms) |
| Visual Cortex | 1.37 | [0,1] | 7.99 | 16.8 | 33.7 | 40.0 |
| Hippo. Culture | 12.67 | [1,2] | 17.90 | 16.8 | 33.7 | 48.0 |

### A. Replication of Synaptic Plasticity in Visual Cortex

The MSQ triplet STDP circuit is used to reproduce the algorithm Nearest-spike model, which is fitted based on the experimental data of the visual cortex slices [4]. The pairing protocol is utilized in this experiment, in which 60 Pre-Post spike pairs with a time interval of $\Delta t = t_{\text{Pre}} - t_{\text{Post}}$ are elicited at different repetition frequencies $\rho$, as shown in Fig. 7(a). It is observed that LTP goes up with $\rho$ increases, which is not possible in the pairwise STDP [4]. The results suggest that the proposed circuit can replicate the algorithmic model accurately, with a NMSE of only 1.78%.

### B. Replication of Synaptic Plasticity in Hippocampal Cultures

The MSQ triplet STDP circuit can also achieve a good match with the algorithmic model that is fitted from experimental data of the hippocampal cultures [4]. Firstly, the time interval between adjacent spike pairs is set to be 1 s through setting $\rho = 1$ Hz, which is long enough to make the effects between pairs negligible. The weight change induced by 60 continuous Pre-Post spike pairs with a delay of $\Delta t$ is shown in Fig. 7(b). By this way, the pairwise STDP learning window is reproduced, which proves that the proposed circuit is compatible with pairwise STDP.

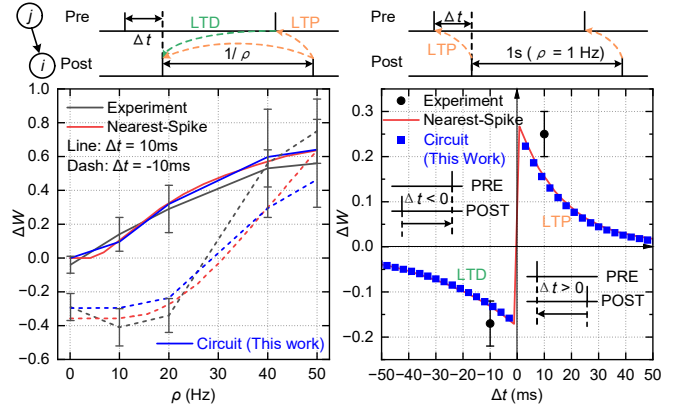

Fig. 7. Replication of the pairwise experiment data in (a) visual cortex with varying repetition frequency $\rho$, and (b) hippocampal cultures with varying time interval $\Delta t$.

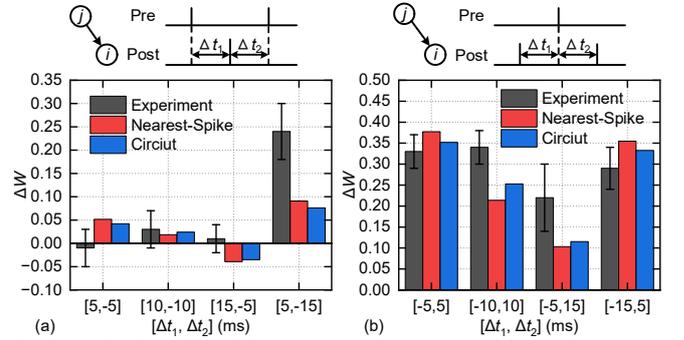

Fig. 8. Replication of the triplet hippocampal cultures experiment data.

Triplet protocol is another experimental method studied, where 60 triplets of Pre-Post-Pre and Post-Pre-Post spikes are repeated at a frequency of $\rho = 1$ Hz, as shown in Fig. 8(a) and (b), respectively. Different combinations of time intervals $\Delta t_1$ and $\Delta t_2$ are used to estimate the weight modulations under certain circumstances. The NMSE between the results of the circuit and algorithm model is 6.48%, which is mainly caused by the time scale of the TDM frame. A frame lasts for 3 ms, which is the minimum time unit of signals. Consequently, the time intervals $\Delta t_1$ and $\Delta t_2$ used in the simulation are approximate values. For example, 10 ms is assumed to be 3 frames which is 9 ms in fact. Additionally, the trace drop within a frame can also deteriorate the accuracy of weight modulation.

Finally, the quadruplet protocol is studied with 60 quadruplets repeated at frequency $\rho = 1$ Hz. In each quadruplet, a Pre-Post pair with a delay of 5 ms is followed by a Post-Pre pair with a delay of -5 ms a time $T$, as shown in Fig. 9. When T is negative, the opposite happens, where the quadruplet becomes Post-Pre-Pre-Post. The results indicate that the proposed circuit is capable of the quadruplet protocol.

TABLE III
Comparison of Capability, Performance, and Components Between This Work and Prior Design

| Designs | Circuit Type | Synapse | STDP rule | Capability | | | Triplet fitting NMSE | Weight modulation signal | Components of synaptic weight update module |
|---|---|---|---|---|---|---|---|---|---|
| | | | | Pairwise | Triplet | Quadruplet | | | |
| [14] | Analog | CMOS | Pairwise | √ | × | × | × | Square wave | 1 OPA, and 6 transistors |
| [15] | Analog | CMOS | Pairwise | √ | × | × | × | Square wave | 5 OPAs, and 8 transistors |
| [6] | Analog | Memristive | Triplet | √ | √ | Not given | 38% | Biomimetic | Not given |
| [7] | Analog | Memristive | Triplet | √ | √ | Not given | 8.68% | Biomimetic | Not given |
| This | Mixed | Memristive | Triplet | √ | √ | √ | 6.48% | Square wave | 9 OPAs, and 25 transistors |

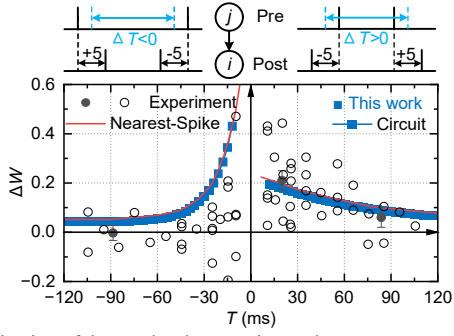

Fig. 9. Replication of the quadruplet experiment data.

*C. Comparison with the State-of-Art Designs*

The comparison between the inhouse design and the state-of-art designs is carried out in this subsection. Compared with these pairwise-STDP circuit designs in [14] and [15] and the prior triplet-STDP circuits in [6] and [7], the advantages of inhouse circuit design scheme can be summarized into threes aspects, they are the capability, performance, and complexity, as listed in TABLE Ⅲ. First, in terms of capabilities, the inhouse design in capable of implementing pair-, triplet-, and quadruplet-STDP algorithms while these in [14] and [15] can only realize the pairwise behavior due to the intrinsic limitation of pairwise-STDP [4]. Second, in the view of performance, the robustness and applicability are enhanced by applying the pulse-width-encoded weight modulation signals, that is, square-wave pulses with identical amplitude but different pulse widths. Comparatively, the utilization of biomimetic spikes in [6] and [7] makes it hard to design the signal generator circuits and control the quantity of weight modulation. Moreover, the inhouse design is more biomimetic in comparison with those in [6] and [7], facilitating its application in implantable medical devices. Third, from the perspective of circuit complexity, all the circuit blocks used in the inhouse design are general modules and each module can be adjusted according to actual needs and available processes. Although, more components are required in comparison with the circuit in [14] and [15], the functions realized by the inhouse circuit are more complex, as mentioned above. In addition, the inhouse circuit is clock-driven, which improves compatibility with conventional digital circuits and enhances its potential as an edge computing device.

## V. Conclusions

In this brief, a mixed-signal circuit scheme called multiple-step quantized (MSQ) triplet STDP is proposed, aiming to implement triplet STDP in memristive synapse-based SNN circuits in a more flexible, robust, and practical way. Multiple RC circuits are employed to characterize the dynamic of each trace individually, which facilitate the implementation of complex spike relationship in the circuit. TDM is utilized to eliminate the interference between spike transmission and weight modulation. The weight modulation signal is encoded into pulse width signals through PWM, which is more practical in actual applications, compared to the complex signals used in related works. By replicating the weight modulation results of the algorithmic model following the pairing, triplet, and quadruplet protocols through simulation, the triplet STDP capability of the circuit is verified.